\documentclass[aps,prd,preprint,superscriptaddress,amsmath,amssymb,showpacs]{revtex4-1}
\usepackage{dcolumn}
\usepackage{graphicx}
\usepackage{float}
\usepackage{physics}
\usepackage[colorlinks=true,allcolors=blue]{hyperref}

\begin{document}

\title{Holographic entanglement entropy under external magnetic field from the EMD model}

\author{Man-Man Sun}
\email{sunmm@zknu.edu.cn}
\affiliation{School of Physics and Telecommunications Engineering, Zhoukou Normal University, Zhoukou 466001, China}

\author{Man-Li Tian}
\email{20252018@zknu.edu.cn}
\affiliation{University Clinic, Zhoukou Normal University, Zhoukou 466001, China}

\author{Zhou-Run Zhu}
\email{zhuzhourun@zknu.edu.cn}
\affiliation{School of Physics and Telecommunications Engineering, Zhoukou Normal University, Zhoukou 466001, China}

\begin{abstract}
In this work, we explore holographic entanglement entropy in the QCD phase diagram under an external magnetic field using an Einstein-Maxwell-dilaton model. We consider both the specious-confinement and deconfined phases. In the perpendicular magnetic field orientation, the strip length shows three distinct branches, and the entanglement entropy develops a swallow-tail structure, indicating a transition between connected and disconnected entanglement surfaces. For the parallel orientation, the behavior is monotonic and no transition occurs. In addition, the difference in entanglement entropy changes smoothly with temperature at small chemical potential, but becomes multivalued at large chemical potential. Increasing the magnetic field restores single-valued behavior. These results are consistent with the black hole thermodynamics and the QCD phase diagram. Our findings show that entanglement entropy can serve as an effective probe of the QCD phase transition.
\end{abstract}

\maketitle

\section{Introduction}\label{sec:01_intro}

Quantum entanglement has become a central concept in the study of strongly correlated quantum systems, providing deep insights into topics that range from quantum information to quantum gravity \cite{Amico:2007ag,Eisert:2008ur}. In quantum chromodynamics (QCD), entanglement measures offer a useful way to investigate the internal structure of strongly interacting matter, especially in the context of the confinement-deconfinement phase transition \cite{Buividovich:2008kq,Nishioka:2006gr,Levin:2006zz,Kitaev:2005dm}. The gauge/gravity duality \cite{Maldacena:1997re,Witten:1998qj,Gubser:1998bc} has greatly advanced this by enabling the computation of entanglement entropy in strongly coupled gauge theories through geometric prescriptions in the dual gravitational theory.

The Ryu-Takayanagi formula \cite{Ryu:2006bv,Ryu:2006ef} connects the entanglement entropy of a boundary subregion in a field theory to the area of a minimal surface that extends into the bulk and ends on the boundary. In confining gauge theories, the holographic entanglement entropy shows a clear transition as the subsystem size varies. For small regions, the entropy scales with the number of colors, which signals the confinement phase \cite{Klebanov:2007ws,Dudal:2016joz,Fujita:2008zv,Kol:2014nqa,Lewkowycz:2012mw}. Such behavior has been seen in both holographic \cite{Cai:2012xh,Braga:2017fsb,Albash:2012pd,Ghodrati:2015rta,Knaute:2017lll} and lattice QCD studies \cite{Buividovich:2008gq,Itou:2015cyu,Katz:2016azl,Bonati:2018uwh}, indicating that entanglement entropy is sensitive to non-perturbative effects.

The holographic approach to entanglement entropy has been extensively developed in various physical backgrounds. Early foundational work established the connection between entanglement entropy and minimal surfaces in AdS space \cite{Ryu:2006bv,Ryu:2006ef,Hubeny:2007xt}. Later studies extended this to confining backgrounds, where a characteristic phase transition in entanglement entropy was observed \cite{Klebanov:2007ws,Dudal:2016joz,Cai:2012sk}. In thermal field theories, holographic entanglement entropy was found to change monotonically with temperature \cite{Yang:2023wuw}. The time evolution of entanglement during thermalization was examined in \cite{Liu:2013iza}. In addition, the complexity of holographic states has been connected to entanglement features through the complexity=volume and complexity=action conjectures \cite{Brown:2015bva,Stanford:2014jda}.

A particularly interesting direction involves entanglement entropy near phase transitions. In holographic QCD models, it acts as a sensitive indicator of the deconfinement transition \cite{Asadi:2022mvo,Zhang:2016rcm,Ali-Akbari:2017vtb}. Studies in holographic QCD have shown how entanglement entropy captures the transition. The behavior of entanglement entropy in the QCD phase diagram has revealed non-monotonic features that may signal the presence of critical points \cite{Li:2020pgn,Dudal:2018ztm}. A key challenge in QCD research is to understand the phase structure under extreme conditions¡ªnamely high temperature, finite chemical potential, and magnetic fields. Strong magnetic fields are known to appear in the early universe and in non-central heavy-ion collisions \cite{Skokov:2009qp,Tuchin:2013ie,Duncan:1992hi,Grasso:2000wj}. The magnetic field breaks spatial isotropy and have a notable impact on QCD properties, giving rise to phenomena like inverse magnetic catalysis \cite{Bali:2011qj,Bruckmann:2013oba,Bohra:2019ebj,Ayala:2014iba,Zhu:2023aaq} and anisotropic string tensions \cite{Bonati:2014ksa,Bonati:2016kxj}. Holographic models that include magnetic fields have been able to reproduce many of these effects \cite{Gursoy:2017wzz,Dudal:2014jfa,Jokela:2013qya,Critelli:2016cvq,Li:2016gfn}, providing a useful framework for studying anisotropic strongly coupled matter. The influence of chemical potential or magnetic field on holographic entanglement entropy has also been examined in \cite{Jain:2022hxl,Arefeva:2020uec}.

The investigation of entanglement entropy across the QCD phase diagram under the combined influence of magnetic fields and chemical potential remains incomplete. In this work, we aim to study the entanglement entropy in the QCD phase diagram with an external magnetic field and finite chemical potential. The transition between small and large black holes can be interpreted as a specious-confinement-deconfinement phase transition \cite{Dudal:2017max}. We employ an Einstein-Maxwell-dilaton (EMD) model to explore the entanglement entropy in the specious-confinement (small black hole) and deconfinement (large black hole) phases. Our goal is to explore how temperature, chemical potential, and magnetic field affect entanglement entropy, with a particular focus on its behavior near the specious-confinement-deconfinement transition.

The paper is organized as follows. In Sec.~\ref{sec:02}, we review the EMD model. In Sec.~\ref{sec:03}, we discuss the holographic entanglement entropy. In Sec.~\ref{sec:04}, we make the conclusion and discussion.

\section{Background geometry}\label{sec:02}
In this section, we provide a brief review of the background geometry for the holographic QCD model at finite chemical potential and magnetic field, as presented in \cite{Bohra:2019ebj}. The corresponding action of the Einstein-Maxwell-dilaton gravity background takes the form
\begin{equation}
\begin{split}
\label{eqa}
 \ S = -\frac{1}{16\pi G_5 }\int d^5 x \sqrt{- g}[R-\frac{f_{1}(\phi)}{4}F_{(1)MN}F^{MN}-\frac{f_{2}(\phi)}{4}F_{(2)MN}F^{MN}-\frac{1}{2}\partial_M \phi \partial^M \phi-V(\phi) ],
 \end{split}
\end{equation}
where $\phi$ denotes the dilaton field and $V(\phi)$ is the potential. $F_{(1)MN}$ and $F_{(2)MN}$ denote the field strength tensors of two U(1) gauge fields. The first gauge field is dual to a neutral flavor current responsible for meson production, while the second is dual to the electromagnetic current. The baryon chemical potential is introduced via the boundary value of the temporal component of the first Abelian gauge field, $A_{(1)M}=A_t (z)\delta^t_M$. The gauge kinetic functions $f_{1}(\phi)$ and $f_{2}(\phi)$ characterize the couplings between the dilaton field and the two U(1) gauge fields. $G_5$ denotes the five-dimensional Newton constant.

The metric $Ans\ddot{a}tze$ adopted from \cite{Bohra:2019ebj} is given by
\begin{equation}
\label{eqb}
\ ds^{2}=\frac{L^2 S(z)}{z^2}[-g(z)dt^2+dx_{1}^{2}+e^{B^2 z^2}(dx_{2}^{2}+dx_{3}^{2})+\frac{dz^{2}}{g(z)}],
\end{equation}
where $S(z)$, $g(z)$, and $L$ represent the scale factor, blackening function, and AdS length scale, respectively. A background magnetic field $B$ is introduced through the second gauge field via $F_{(2)MN} = Bdx_2 \wedge dx_3$, which breaks the $SO(3)$ rotational symmetry as $B$ is aligned along the $x_1$-direction. In this metric, $B$ is a five-dimensional field with mass dimension one; the corresponding physical four-dimensional magnetic field $\mathfrak{B}$, of mass dimension two, is obtained via $\mathfrak{B}\sim B/L$ \cite{DHoker:2009ixq,Dudal:2015wfn}. As we are concerned only with qualitative effects, we adopt the five-dimensional field $B$ throughout this work.

The scale factor $S(z)$ is expressed as \cite{Bohra:2019ebj}
\begin{equation}
 \label{eqj}
\begin{split}
 & S(z) = e^{2P(z)}.
 \end{split}
\end{equation}

The blackening function $g(z)$ is given by \cite{Bohra:2019ebj}
\begin{equation}
 \label{eqk}
 g(z) = 1+\int^{z}_0 d\xi \xi^3 e^{-B^2 \xi^2-3P(\xi)}[K_1 + \frac{\widetilde{\mu}^2}{2c L^2}e^{c \xi^2}],
\end{equation}
with
\begin{equation}
 \label{eql}
 K_1 = -\frac{[1+ \frac{\widetilde{\mu}^2}{2c L^2}\int^{z_h}_0 d\xi \xi^3 e^{-B^2 \xi^2-3P(\xi)+c\xi^2}]}{\int^{z_h}_0 d\xi \xi^3 e^{-B^2 \xi^2-3P(\xi)}}.
\end{equation}

The dilaton field expressed in terms of $P(z)$ reads \cite{Bohra:2019ebj}
\begin{equation}
 \label{eqm}
 \phi(z)= \int dz \sqrt{-\frac{2}{z} \left(3 z P''(z)-3zP'(z)^2 +6P'(z)+2 B^4 z^3+2B^2 z \right) }+K_2.
\end{equation}
where $K_2$ ensures that $\phi$ vanishes near the asymptotic boundary.

The dilaton potential is given by \cite{Bohra:2019ebj}
\begin{equation}
\label{eqn}
\begin{split}
\ & V(z)=\frac{g(z)}{L^2}\left(-\frac{9B^2 z^3 S'(z)}{2S(z)^2}+\frac{10B^2 z^2}{S(z)} -\frac{3z^2 S'(z)^2 }{S(z)^3}+\frac{12z S'(z)}{S(z)^2}+\frac{z^2 \phi'(z)^2}{2S(z)}-\frac{12}{S(z)}\right) \\
& -\frac{z^4 f_{1}(z)A'_{t}(z)^2}{2L^4 S(z)^2}+\frac{g'(z)}{L^2}\left(-\frac{B^2 z^3}{S(z)}-\frac{3z^2 S'(z)}{2S(z)^2}+\frac{3z}{S(z)} \right).
\end{split}
\end{equation}

The Hawking temperature and entropy in this background are \cite{Bohra:2019ebj}
\begin{equation}
\label{eqo}
\begin{split}
& T=-\frac{z^3_h e^{-3P(z_h)-B^2 z^2_h}}{4\pi}[K_1 + \frac{\widetilde{\mu}^2}{2c L^2}e^{c z^2_h}],\\
& S= \frac{e^{B^2 z^2_h +3P(z_h)}}{4 z^3_h},
 \end{split}
\end{equation}
where the Newton constant $G_5$ is set to unity.

In Ref.\cite{Zhu:2023aaq}, the authors choose the following the form of $P(z)$ in Ref. \cite{Li:2017tdz}
\begin{equation}
\label{eqp}
 \ P(z)= -a\log(b z^2 +1).
\end{equation}

In contrast to \cite{Bohra:2019ebj}, the work of Ref.\cite{Zhu:2023aaq} focuses on the phase structure of QCD for light quarks. As discussed in Ref.\cite{Zhu:2023aaq}, this holographic model captures a number of desirable anisotropic QCD features, as summarized below.

The model parameters are fixed by comparison with QCD observables. The parameter \(c\) is set to \(0.227\) at \(B = 0\) by fitting the \(\rho\) meson mass spectrum \cite{Yang:2014bqa}. To reproduce the confinement-deconfinement transition temperature at \(B = 0\) and \(\mu = 0\), one can fix \(a = 3.943\) and \(b = 0.0158\) \cite{Zhu:2023aaq}. Then one can find the phase transition temperature is around \(150\)-\(160\) MeV and is consistent with lattice QCD. The consistency of the model is carefully checked. The dilaton field satisfies \(\phi(0) = 0\), ensuring an asymptotically AdS spacetime. The gauge kinetic functions remain positive throughout the bulk, satisfying the null energy condition. The dependence of the dilaton potential $V(\phi)$ on the thermodynamic parameters $B$ and $\mu$ has been carefully examined\cite{Zhu:2023aaq}. The curves of $V(\phi)$ for different values of $B$, $\mu$, and $T$ almost coincide, indicating that such dependence is negligible. This justifies, within the approximation scheme, treating the reconstructed model as effectively describing a single dual theory across different thermodynamic states. The Gubser stability criterion \cite{Gubser:2008px} requires that the dilaton potential remains bounded from below along the holographic flow, thereby excluding tachyonic instabilities. The authors of Ref.\cite{Zhu:2023aaq} have verified that our reconstructed $V(z)$ satisfies this condition.

Analysis of the black hole thermodynamics reveals that the chemical potential drives the crossover toward a first-order transition, whereas the magnetic field reverses this effect, converting the first-order transition back to a crossover\cite{Zhu:2023aaq}. The resulting phase diagrams in the \(T\)-\(\mu\) and \(T\)-\(B\) planes exhibit inverse magnetic catalysis. The authors of Ref.\cite{Zhu:2023aaq} also examine the pressure, baryon density, entropy density, specific heat, and sound speed near the phase transition temperature. The nonmonotonic and nontrivial behaviors observed in these quantities are used to characterize the nature of the phase transition, but no quantitative comparison to lattice QCD data is attempted. Our results of holographic entanglement should therefore be interpreted as qualitative and model-dependent.

\section{Holographic entanglement entropy in the Einstein-Maxwell-dilaton gravity background}\label{sec:03}

In this section, we calculate the entanglement entropy in the holographic EMD model. The entanglement entropy \( S_A \) can be computed using the following formula from holography \cite{Ryu:2006bv}
\begin{eqnarray}
S^{EE}=\frac{\text{Area}(\gamma_A)}{4 G_5},
\label{RT}
\end{eqnarray}
where \(\gamma_A\) denotes a three-dimensional minimal area surface in the asymptotically AdS\(_5\) spacetime. This geometric prescription reproduces the entanglement entropy in two-dimensional conformal field theories, thereby offering a reliable method for evaluating entanglement entropy in strongly coupled gauge theories. Our calculations are performed in the Einstein frame. Switching to the string frame does not alter the results, provided that an extra dilaton-dependent exponential factor is included in the prescription (\ref{RT}), following \cite{Ryu:2006ef}.

The above equation can also be written in the following way
\begin{equation}
S(A) = \frac{1}{4G_{5}}\int d^{3}x \sqrt{g_{ind}}\ ,
\label{heestring}
\end{equation}
where $g_{ind}$ denotes the induced metric on \(\gamma_A\).

We rewrite the metric (\ref{eqb})
\begin{equation}
\label{eqbb4}
\ ds^{2}= g_{tt} dt^2 + g_{xx_{1}} dx_{1}^{2}+ g_{xx_{2}}(dx_{2}^{2}+dx_{3}^{2})+g_{zz} dz^{2},
\end{equation}
where $g_{tt}= -\frac{L^2 e^{2P(z)}}{z^2} g(z)$, $g_{xx_{1}}=\frac{L^2 e^{2P(z)}}{z^2}$, $g_{xx_{2}}=\frac{L^2 e^{2P(z)}}{z^2} e^{B^2 z^2}$  and $g_{zz}=\frac{L^2 e^{2P(z)}}{z^2} \frac{1}{g(z)}$.

In the presence of a background magnetic field, the entangling surface can align parallel and perpendicular to the magnetic field. To be specific, one can consider the boundary subsystem with the domain $-x_{\parallel}/2\leq x_1 \leq x_{\parallel}/2$ for parallel case. As for perpendicular case, the strip subsystem with the domain $-x_{\perp}/2\leq x_2 \leq x_{\perp}/2$. Here, $x$ represents the strip length. The strip subsystem admits connected and disconnected surfaces. The formulas for the strip length and the entanglement entropy of the connected and disconnected surfaces are obtained by extremizing the area functional for a strip-shaped subsystem. For a given orientation of the magnetic field, the induced metric on the minimal surface is computed. The details follow the standard holographic prescription \cite{Ryu:2006bv,Ryu:2006ef,Hubeny:2007xt}

In parallel case, the strip length for the connected surface is
\begin{equation}
\label{eq14}
\begin{split}
x_{\parallel}=2\int_{0}^{z_{c}} dz ( \frac{g_{xx_{1}}(z)}{g_{zz}(z)}(\frac{g_{xx_{1}}(z)g_{xx_{2}}^2(z)}{g_{xx_{1}}(z_{c})g_{xx_{2}}^2(z_{c})}-1  ) )^{-\frac{1}{2}},
 \end{split}
\end{equation}
where $z_{c}$ represents the turning point of the minimal area surface.

In parallel case, the entanglement entropy for the connected surface is
\begin{equation}
\label{eq15}
\begin{split}
S_{con,\parallel}= \frac{V_2}{2G_5} \int_{0}^{z_{c}} dz \frac{\sqrt{g_{xx_{1}}(z)g_{zz}(z)} g_{xx_{2}}^2(z) } {\sqrt{g_{xx_{1}}(z) g_{xx_{2}}^2(z)-g_{xx_{1}}(z_c) g_{xx_{2}}^2(z_c)}  } ,
 \end{split}
\end{equation}
where $V_2$ denotes the area of the surface in the $(x_2, x_3)$ plane.

In parallel case, the entanglement entropy for the disconnected surface is
\begin{equation}
\label{eq16}
\begin{split}
S_{dis,\parallel}= \frac{V_2}{2G_5} \int_{0}^{z_{h}} dz \sqrt{g_{zz}(z)g_{xx_{2}}^2(z)} +\frac{1}{2}x_{\parallel}\sqrt{g_{xx_{1}}(z_h)g_{xx_{2}}^2(z_h)}.
 \end{split}
\end{equation}

In perpendicular case, the strip length for the connected surface is
\begin{equation}
\label{eq17}
\begin{split}
x_{\perp}=2\int_{0}^{z_{c}} dz ( \frac{g_{xx_{2}}(z)}{g_{zz}(z)}(\frac{g_{xx_{1}}(z)g_{xx_{2}}^2(z)}{g_{xx_{1}}(z_{c})g_{xx_{2}}^2(z_{c})}-1  ) )^{-\frac{1}{2}},
 \end{split}
\end{equation}
where $z_{c}$ represents the turning point of the minimal area surface.

In perpendicular case, the entanglement entropy for the connected surface is
\begin{equation}
\label{eq18}
\begin{split}
S_{con,\perp}= \frac{V_2}{2G_5} \int_{0}^{z_{c}} dz \frac{\sqrt{g_{zz}(z)}g_{xx_{1}}(z) g_{xx_{2}}^\frac{3}{2}(z) } {\sqrt{g_{xx_{1}}(z) g_{xx_{2}}^2(z)-g_{xx_{1}}(z_c) g_{xx_{2}}^2(z_c)}  } ,
 \end{split}
\end{equation}
where $V_2$ denotes the area of the surface in the $(x_2, x_3)$ plane.

In perpendicular case, the entanglement entropy for the disconnected surface is
\begin{equation}
\label{eq19}
\begin{split}
S_{dis,\perp}= \frac{V_2}{2G_5} \int_{0}^{z_{h}} dz \sqrt{g_{zz}(z)g_{xx_{1}}(z)g_{xx_{2}}(z)} +\frac{1}{2}x_{\perp}\sqrt{g_{xx_{1}}(z_h)g_{xx_{2}}^2(z_h)}.
 \end{split}
\end{equation}

It should be mentioned that the background geometry becomes thermal-AdS when $z_h=\infty$ in Eq.(\ref{eq16}) and (\ref{eq19}). The second term of Eq.(\ref{eq16}) and (\ref{eq19}) becomes zero for the thermal-AdS background. Thus, the entanglement entropy associated with the disconnected surface is independent of $x$ in the thermal-AdS background. In contrast, for the AdS black hole background, this second term yields a nonzero contribution to the entanglement entropy.

One can define $\Delta S^{EE} \equiv S_{\rm con} - S_{\rm dis}$ as the difference between the holographic entanglement entropy of the connected and disconnected surfaces. This determines which surface configuration minimizes the area for a given strip length $x$. A negative $\Delta S^{EE}$ indicates that the connected surface is preferred, while a positive $\Delta S$ signals that the disconnected surface is favored. The sign change of $\Delta S^{EE}$ as $x$ varies marks a phase transition in the preferred minimal surface, which holographically corresponds to the confinement-deconfinement transition \cite{Klebanov:2007ws,Dudal:2016joz}. Although $\Delta S^{EE}$ captures qualitative features similar to black hole thermodynamics, especially the anisotropic dependence on the magnetic field orientation. This orientation dependence is a non-thermodynamic observable that complements the thermodynamic studies in Ref.\cite{Zhu:2023aaq}.

\begin{figure}[H]
    \centering
      \setlength{\abovecaptionskip}{0.1cm}
    \includegraphics[width=16cm]{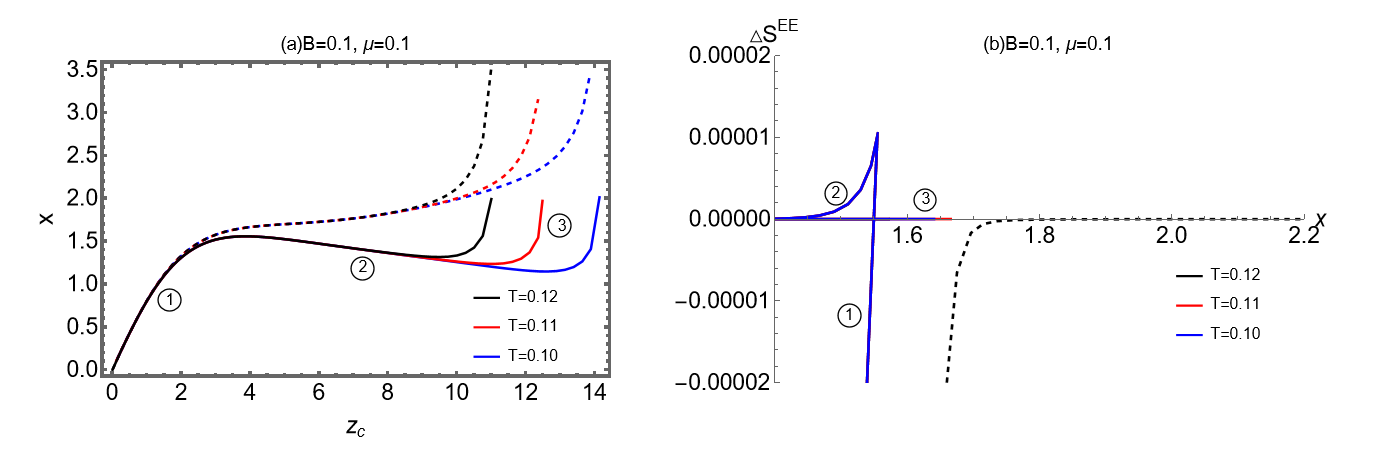}
    \caption{\label{fig1} (a) Strip length $x$ versus $z_c$ at different temperatures in the small black hole. (b) $\Delta S^{EE}= S_{con}-S_{dis}$ as a function of strip length $x$ at different temperatures in the small black hole. The black, blue and red line denote $T = 0.12,\ 0.11,\ 0.1$, respectively. The solid line (dashed line) represents the perpendicular (parallel) case. In units GeV.}
\end{figure}

We emphasize that the simultaneous presence of large chemical potential and strong magnetic field in equilibrium QCD matter is not realized in current heavy ion collision. As discussed in Ref.\cite{Bali:2011qj}, the magnetic fields decay rapidly during the pre-equilibrium stage. Therefore, the present study is a theoretical exploration rather than a direct prediction for experimental observables. The five-dimensional magnetic field $B$ has mass dimension one, as it appears in the combination $e^{B^2 z^2}$ with the holographic coordinate $z$. The physical four-dimensional magnetic field $\mathfrak{B}$ is related to the five-dimensional field by $\mathfrak{B} \sim B/L$, where $L$ is the AdS length scale, as discussed in Refs.\cite{DHoker:2009ixq,Dudal:2015wfn}. Since we set $L=1$ throughout our numerical calculations and label the five-dimensional field $B$ in units of GeV.

We perform a holographic study of entanglement entropy in both the specious-confinement (small black hole) and deconfinement (large black hole) phases. Fig.~\ref{fig1} shows how the strip length and entanglement entropy behave at different temperatures within the small black hole background. In the perpendicular case, Fig.~\ref{fig1}(a) reveals a clear distinction in the strip length behavior. The relation between $x$ and $z_c$ splits into three distinct branches, labeled \textcircled{1}, \textcircled{2}, and \textcircled{3}. Along the first branch (\textcircled{1}), $x$ increases with $z_c$ until it hits a maximum $x_{max}$. Then, along the second branch (\textcircled{2}), $x$ drops from $x_{max}$ down to a minimum $x_{min}$. Finally, on the third branch (\textcircled{3}), $x$ begins to rise again from $x_{min}$. The presence of these three branches makes the transition between different entangling surfaces more involved. For the parallel case, $x$ simply increases monotonically with $z_c$.

Fig.~\ref{fig1}(b) presents the difference in entanglement entropy between connected and disconnected surfaces within the specious-confinement phase. In the perpendicular case, a swallow-tail structure \textcircled{2} appears. This branch consistently yields higher entanglement entropy than branches \textcircled{1} and \textcircled{3}, suggesting that it corresponds to a saddle point of the minimal area functional. As the strip length $x$ increases, the system transitions from a connected to a disconnected surface. For the parallel case, however, no such phase transition occurs. Instead, the disconnected surface always gives a larger entanglement entropy than the connected one.

\begin{figure}[H]
    \centering
      \setlength{\abovecaptionskip}{0.1cm}
    \includegraphics[width=16cm]{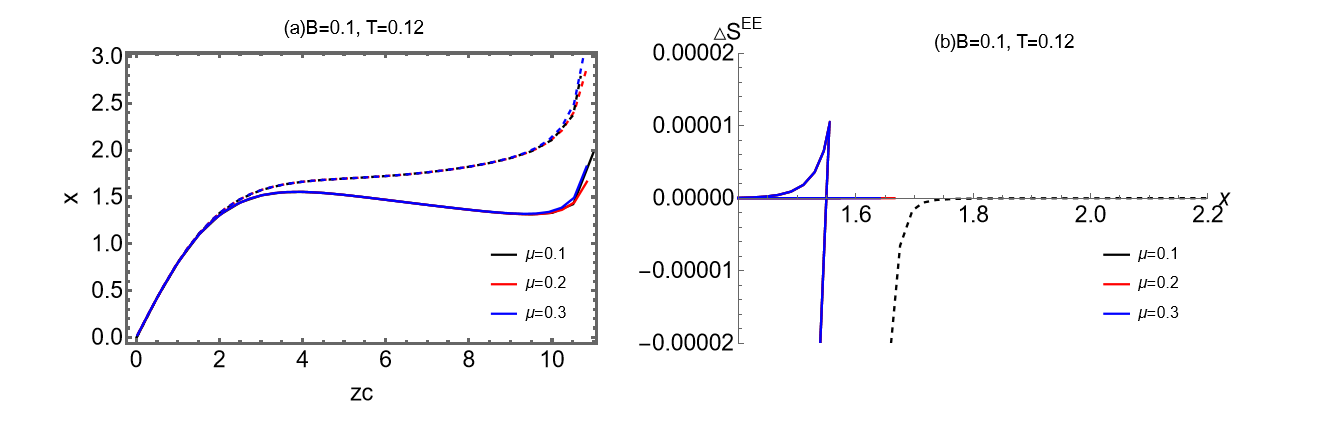}
    \caption{\label{fig2} (a) Strip length $x$ versus $z_c$ at different chemical potential in the small black hole. (b) $\Delta S^{EE}= S_{con}-S_{dis}$ as a function of strip length $x$ at different chemical potential in the small black hole. The black, blue and red line denote $\mu = 0.1,\ 0.2,\ 0.3$, respectively. The solid line (dashed line) represents the perpendicular (parallel) case. In units GeV.}
\end{figure}

\begin{figure}[H]
    \centering
      \setlength{\abovecaptionskip}{0.1cm}
    \includegraphics[width=9cm]{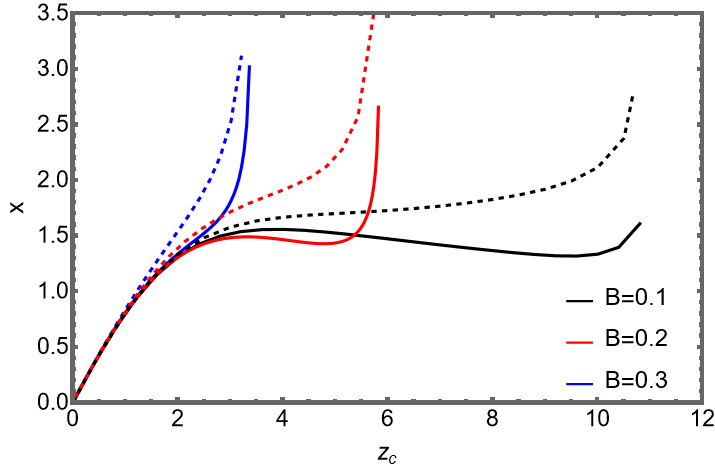}
    \caption{\label{fig3} Strip length $x$ versus $z_c$ at different magnetic field in the small black hole. The black, blue and red line denote $B = 0.1,\ 0.2,\ 0.3$, respectively. The solid line (dashed line) represents the perpendicular (parallel) case. In units GeV.}
\end{figure}

The behavior of entanglement entropy under chemical potential is similar to that under temperature. In Fig.~\ref{fig2}, we show the strip length and entanglement entropy at various chemical potential within the small black hole background. From Fig.~\ref{fig2}(a) and (b), one can see that in the perpendicular case, both the strip length and entanglement entropy exhibit three distinct branches, and a phase transition occurs between connected and disconnected surfaces. In contrast, no such transition appears in the parallel case. It is worth noting that the temperature dependence of entanglement entropy at $\mu=0$ in the small black hole has already been studied in Ref.~\cite{Dudal:2018ztm}. In the present work, we extend the analysis to include finite chemical potential and magnetic field.

\begin{figure}[H]
    \centering
      \setlength{\abovecaptionskip}{0.1cm}
    \includegraphics[width=16cm]{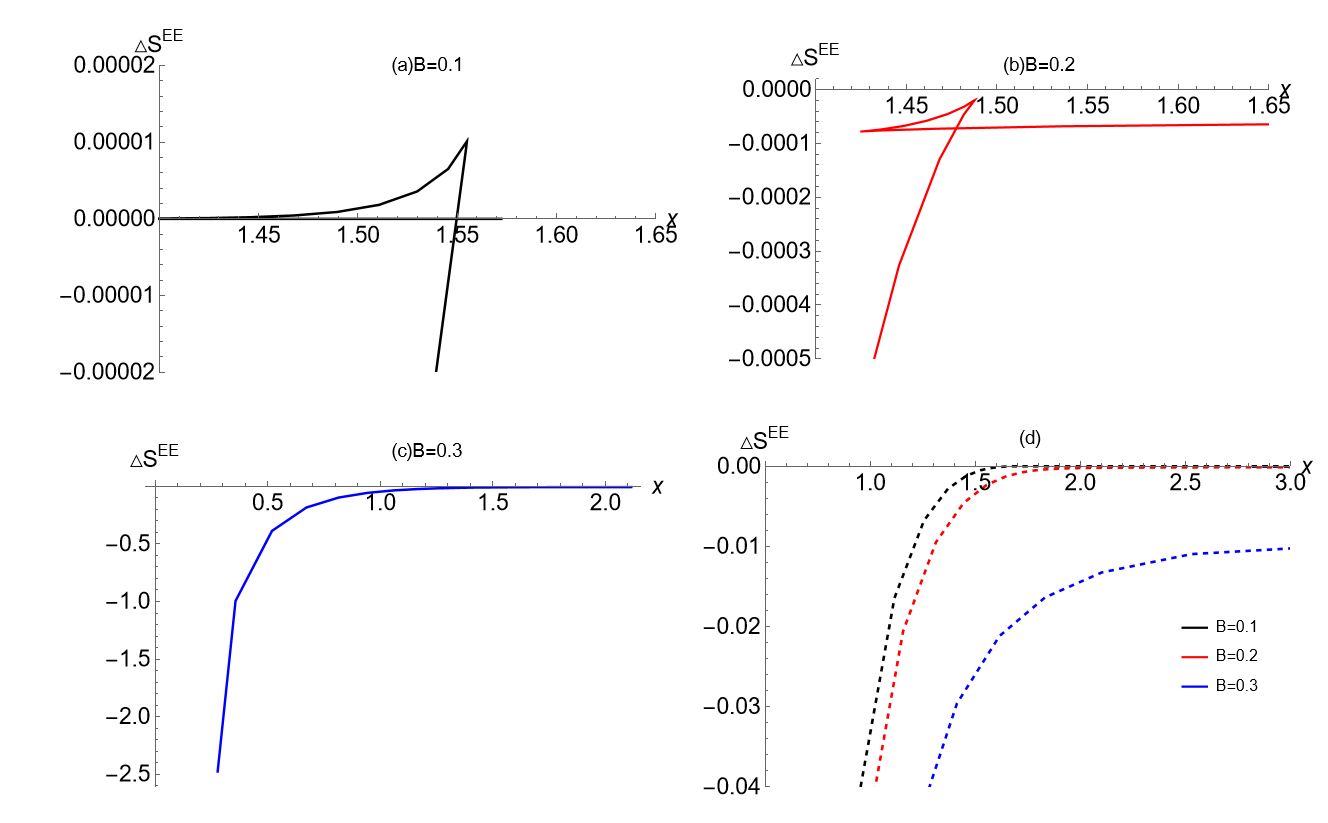}
    \caption{\label{fig4} $\Delta S^{EE}= S_{con}-S_{dis}$ as a function of strip length $x$ at different magnetic field in the small black hole. The black, blue and red line denote $B = 0.1,\ 0.2,\ 0.3$, respectively. The solid line (dashed line) represents the perpendicular (parallel) case. In units GeV.}
\end{figure}

Fig.~\ref{fig3} illustrates how the strip length behaves under different magnetic field strengths in the small black hole background. In the perpendicular case, three distinct branches appear, and a phase transition occurs between connected and disconnected surfaces. As the magnetic field grows, branch \textcircled{2} gradually fades away. For the parallel case, $x$ increases monotonically with $z_c$.

Fig.~\ref{fig4} shows the difference in entanglement entropy under varying magnetic field. When the field is small, a swallow-tail structure labeled \textcircled{2} appears in the perpendicular case, signaling a transition from a connected surface to a disconnected one. As the magnetic field strengthens, branch \textcircled{2} disappears, and the entanglement entropy difference becomes a monotonic function of $x$ at large $B$. It is worth noting that the influence of a magnetic field on entanglement entropy has also been examined in Ref.~\cite{Jain:2022hxl}. In the confining phase, the entanglement entropy undergoes a phase transition at a critical strip length, which increases under a parallel magnetic field and decreases under a perpendicular one.

\begin{figure}[H]
    \centering
      \setlength{\abovecaptionskip}{0.1cm}
    \includegraphics[width=16cm]{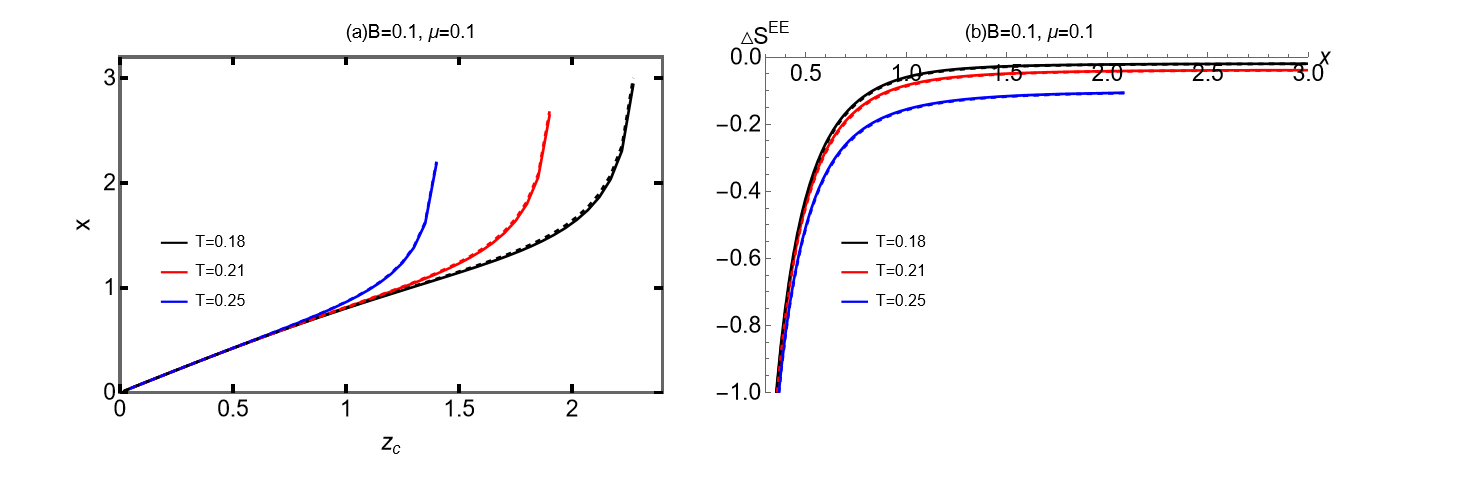}
    \caption{\label{fig5} (a) Strip length $x$ versus $z_c$ at different temperatures in the large black hole. (b) $\Delta S^{EE}= S_{con}-S_{dis}$ as a function of strip length $x$ at different temperatures in the large black hole. The black, blue and red line denote $T = 0.18,\ 0.21,\ 0.25$, respectively. The solid line (dashed line) represents the perpendicular (parallel) case. In units GeV.}
\end{figure}

\begin{figure}[H]
    \centering
      \setlength{\abovecaptionskip}{0.1cm}
    \includegraphics[width=16cm]{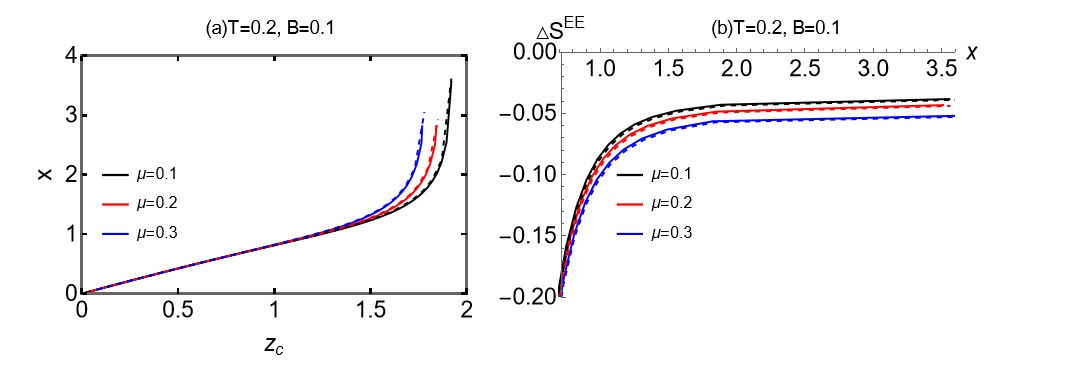}
    \caption{\label{fig6} (a) Strip length $x$ versus $z_c$ at different chemical potential in the large black hole. (b) $\Delta S^{EE}= S_{con}-S_{dis}$ as a function of strip length $x$ at different chemical potential in the large black hole. The black, blue and red line denote $\mu = 0.1,\ 0.2,\ 0.3$, respectively. The solid line (dashed line) represents the perpendicular (parallel) case. In units GeV.}
\end{figure}

\begin{figure}[H]
    \centering
      \setlength{\abovecaptionskip}{0.1cm}
    \includegraphics[width=16cm]{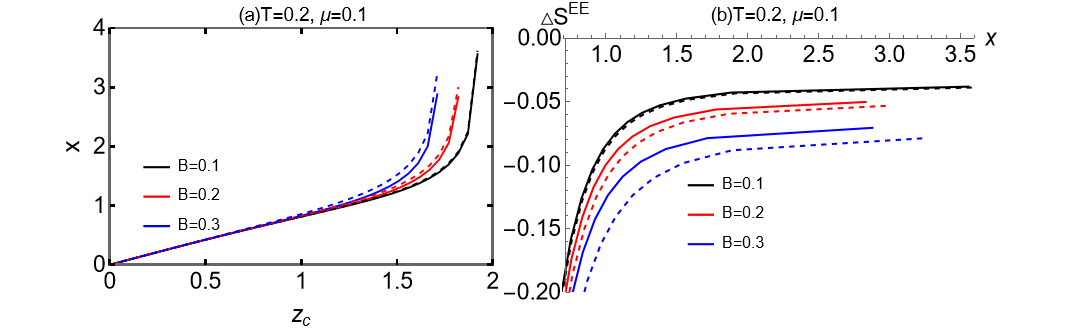}
    \caption{\label{fig7} (a) Strip length $x$ versus $z_c$ at different magnetic field in the large black hole. (b) $\Delta S^{EE}= S_{con}-S_{dis}$ as a function of strip length $x$ at different magnetic field in the large black hole. The black, blue and red line denote $B = 0.1,\ 0.2,\ 0.3$, respectively. The solid line (dashed line) represents the perpendicular (parallel) case. In units GeV.}
\end{figure}

Fig.~\ref{fig5} shows the strip length and entanglement entropy at various temperatures in the large black hole background, which corresponds to the deconfinement phase. In both the perpendicular and parallel cases, $x$ increases monotonically with $z_c$. One also can find that the disconnected surface always gives a larger value than the connected one from the results of entanglement entropy, indicating that no phase transition occurs in this phase.

Figs.~\ref{fig6} and~\ref{fig7} examine how chemical potential and magnetic field affect the strip length and entanglement entropy in the large black hole, respectively. Their effects resemble those of temperature. $x$ grows monotonically with $z_c$ in all cases, and the disconnected surface consistently yields higher entanglement entropy than the connected one.

\begin{figure}[H]
    \centering
      \setlength{\abovecaptionskip}{0.1cm}
    \includegraphics[width=16cm]{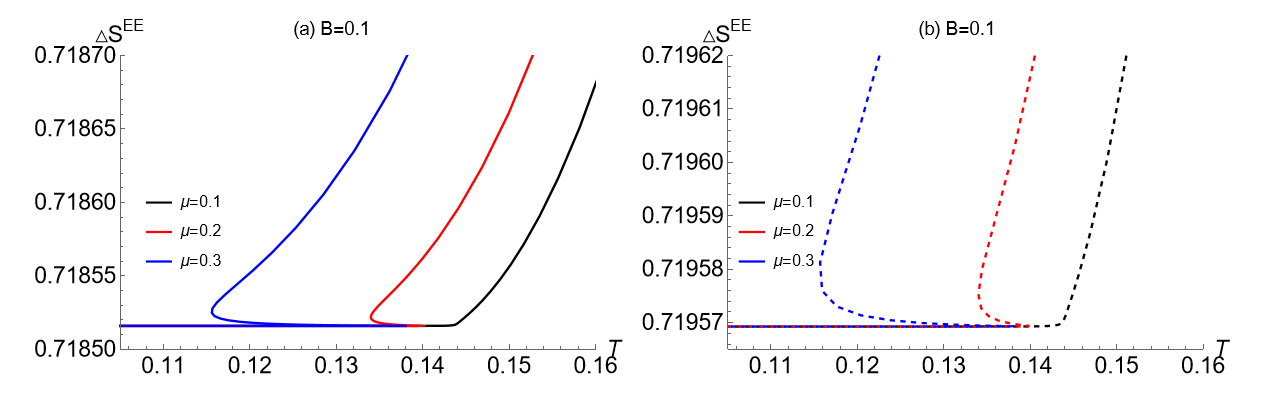}
    \caption{\label{fig8} $\Delta S^{EE}= S_{AdS-BH}-S_{Thermal-AdS}$ as a function of temperature at different chemical potential. (a) for perpendicular case, while (b) for parallel case.}
\end{figure}

\begin{figure}[H]
    \centering
      \setlength{\abovecaptionskip}{0.1cm}
    \includegraphics[width=16cm]{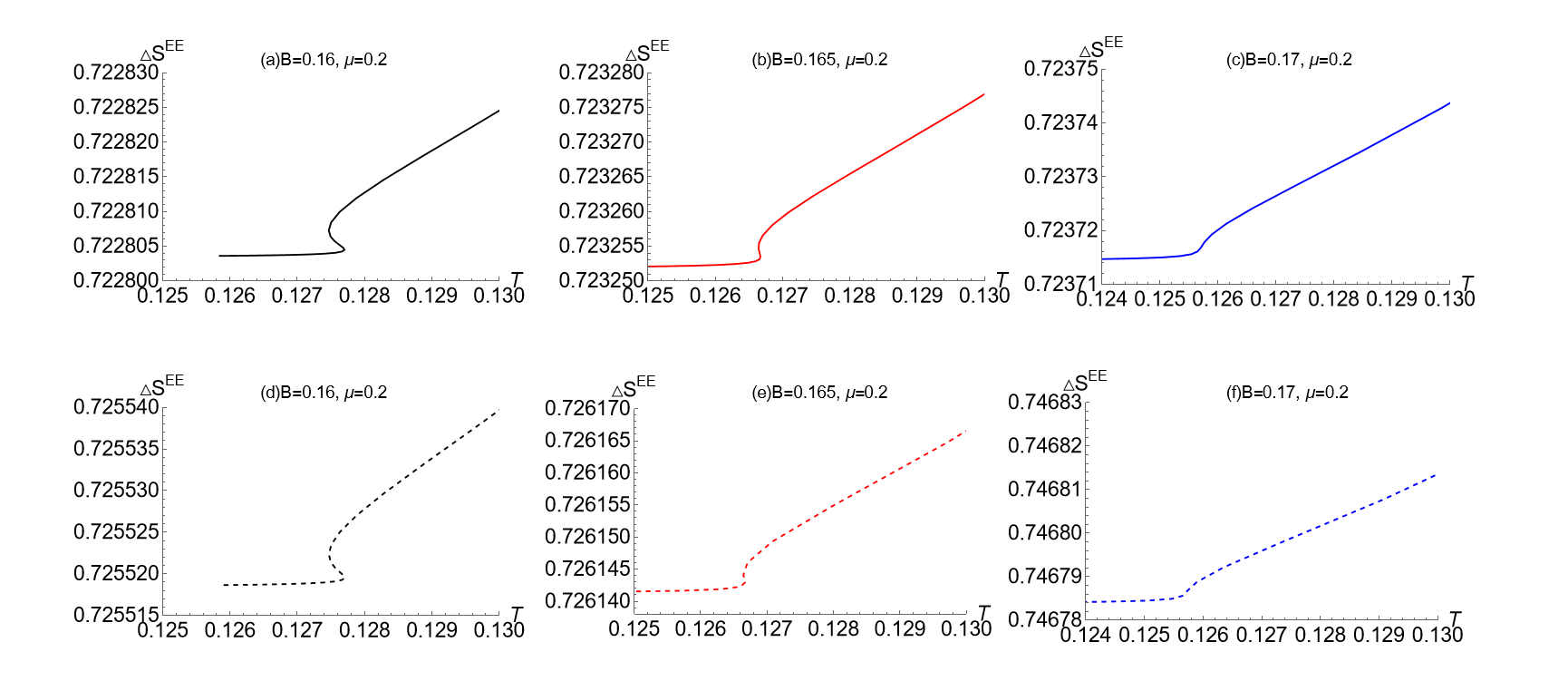}
    \caption{\label{fig9} $\Delta S^{EE}= S_{AdS-BH}-S_{Thermal-AdS}$ as a function of temperature at different magnetic field. (a), (b) and (c) for perpendicular case, while (d), (e) and (f) for parallel case.}
\end{figure}

Fig.~\ref{fig8} presents $\Delta S^{EE}= S_{AdS-BH}-S_{Thermal-AdS}$ as a function of temperature under various chemical potentials. We use $S_{Thermal-AdS}$ to cancel the divergence of $S_{AdS-BH}$ in Fig.~\ref{fig8}. For a fixed strip length $x$, this term is constant and temperature-independent. In Figs.~\ref{fig8} and~\ref{fig9}, we take $x = 0.5  \text{GeV}^{-1}$. As seen in Fig.~\ref{fig8}, $\Delta S^{EE}$ is single-valued at small $\mu$ but becomes multivalued when $\mu$ is large. Fig.~\ref{fig9} shows $\Delta S^{EE}$ as a function of temperature at different magnetic field strengths. One can see that the multivalued structure gradually gives way to single-valued behavior as the magnetic field grows.

We now turn to the connection between entanglement entropy and the QCD phase transition. The entanglement entropy is computed from a minimal surface that stretches from the asymptotic boundary into the bulk, thus it may naturally carry signatures of the black hole phase transition. From holography, the transition between small and large black holes corresponds to the specious-confinement-deconfinement transition in the dual field theory \cite{Dudal:2017max}.

Using the same EMD model, Ref.~\cite{Zhu:2023aaq} explores black hole thermodynamics and the QCD phase diagram. The results show that at small $\mu$, the free energy and entropy of the black hole are single-valued, while at large $\mu$ they become multivalued, indicating a crossover at low chemical potential and a first-order transition at high chemical potential. Namely, increasing $\mu$ turns the crossover into a first-order transition. A sufficiently strong magnetic field restores single-valuedness, meaning that the magnetic field pushes the first-order transition to a crossover. These findings are consistent with Figs.~\ref{fig8} and~\ref{fig9}. The entanglement entropy exhibits single-valued at small $\mu$ and multivalued at large $\mu$, signaling a crossover to first-order transition. The large magnetic field restores single-valued phenomenon, reflecting a return to crossover behavior. This suggests that entanglement entropy is sensitive to the nature of the QCD phase transition.

\section{Conclusion and discussion}\label{sec:04}

In this work, we systematically investigated holographic entanglement entropy across the QCD phase diagram using an Einstein-Maxwell-dilaton model, with particular focus on the effects of temperature, chemical potential, and an external magnetic field. Our study covered both the specious-confinement (small black hole) and deconfinement (large black hole) phases, as well as parallel and perpendicular orientations of the magnetic field.

In the specious-confinement phase with a perpendicular magnetic field, the strip length as a function of the turning point exhibits three distinct branches, leading to a swallow-tail structure in the entanglement entropy difference, which indicates a phase transition between connected and disconnected entangling surfaces. In the parallel case, no such transition occurs, and the disconnected surface always yields larger entanglement entropy. The anisotropic behavior observed here originates from the breaking of rotational invariance by the external magnetic field. The strip length $x$ becomes non-monotonic when the strip is oriented perpendicular to the field, giving rise to multiple extremal surfaces. The resulting swallow-tail structure contains an intermediate branch that corresponds to an unstable saddle point, and the system undergoes a first-order transition between connected and disconnected surfaces as $x$ varies. In the parallel orientation, by contrast, translational invariance along the field direction is preserved. The $x(z_c)$ behaves as only a single, monotonic branch. Consequently, no phase transition occurs, and the disconnected surface is always energetically favored.

The effect of chemical potential resembles that of temperature. At small chemical potential the entanglement entropy difference is single-valued (crossover), while at large chemical potential it becomes multivalued (first-order transition). Increasing the magnetic field gradually suppresses the swallow-tail structure and restores single-valuedness, converting the first-order transition back to a crossover. In the deconfinement phase, no phase transition is observed. Our results show that the sign and behavior of $\Delta S$ reflect the nature of the holographic phase transition. Although $\Delta S$ itself is not a direct measure of entanglement for mixed states, it serves as a useful diagnostic for the transition between connected and disconnected minimal surfaces.

Our results show that entanglement entropy captures the same qualitative features as black hole thermodynamics, establishing it as a reliable and sensitive probe of the specious-confinement-deconfinement phase transition in the QCD phase diagram. This work provids a systematic holographic study of entanglement entropy under the combined influence of magnetic field and chemical potential, deepening our understanding of the QCD phase diagram.

\section*{Acknowledgments}

Manman Sun is supported by the National Natural Science Foundation of China under Grant No. 12305076. Zhou-Run Zhu is supported by the Natural Science Foundation of Henan Province of China under Grant No. 242300420947. Zhou-Run Zhu is also supported by the High Level Talents Research and Startup Foundation Projects for Doctors of Zhoukou Normal University No. ZKNUC2023018.

\end{document}